\documentclass[10pt,conference]{IEEEtran}
\usepackage{cite}
\usepackage[intlimits]{amsmath}
\usepackage{amssymb,amsfonts}
\usepackage{mathtools}
\usepackage{graphicx}
\usepackage{textcomp}
\usepackage{cleveref}
\usepackage{braket}
\ifCLASSOPTIONcompsoc
    \usepackage[listofformat=simple,subrefformat=subsimple,caption=false,font=normalsize,labelfont=sf,textfont=sf]{subfig}
\else
    \usepackage[listofformat=simple,subrefformat=subsimple,caption=false,font=footnotesize]{subfig}
\fi

\usepackage{flushend}
\usepackage{stfloats}

\fnbelowfloat
\crefformat{equation}{(#2#1#3)}
\crefrangeformat{equation}{(#3#1#4) to (#5#2#6)}

\newcommand{\Rplus}{\mathbb{R}_{> 0}}
\newcommand{\expectation}[1]{\mathbb{E}\sqparens{#1}}
\newcommand{\e}{\mathrm{e}}
\newcommand{\diff}{\mathrm{d}}
\renewcommand{\i}{\boldsymbol{\mathrm{i}}}
\newcommand{\N}{\mathbb{N}}
\newcommand{\C}{\mathbb{C}}
\newcommand{\eqequal}[1]{\overset{\cref{#1}}{=}}

\DeclarePairedDelimiter\parens{(}{)}
\DeclarePairedDelimiter\sqparens{[}{]}
\DeclarePairedDelimiter\abs{\lvert}{\rvert}

\newcommand{\indI}{j}
\newcommand{\indII}{l}

\newcommand{\varMaturity}{T}

\newcommand{\varInterest}{r}
\newcommand{\varStrike}{K}
\newcommand{\varLogStrike}{k}
\newcommand{\varLowestLogStrike}{k_0}

\def\BibTeX{{\rm B\kern-.05em{\sc i\kern-.025em b}\kern-.08em
    T\kern-.1667em\lower.7ex\hbox{E}\kern-.125emX}}

\begin{document}

\title{Pricing of European Calls\\
with the\\
Quantum Fourier Transform
}

\author{\IEEEauthorblockN{Tom Ewen}
\IEEEauthorblockA{\textit{Department of Financial Mathematics} \\
\textit{Fraunhofer ITWM}\\
Kaiserslautern, Germany \\
tom.ewen@itwm.fraunhofer.de \\
0009-0007-4028-6698
}
}

\maketitle

\begin{abstract}
    The accurate valuation of financial derivatives plays a pivotal role in the finance industry.
    Although closed formulas for pricing are available for certain models and option types, exemplified by the European Call and Put options in the Black-Scholes Model, the use of either more complex models or more sophisticated options precludes the existence of such formulas, thereby requiring alternative approaches.
    The Monte Carlo simulation, an alternative approach effective in nearly all scenarios, has already been challenged by quantum computing techniques that leverage Amplitude Estimation.
    Despite its theoretical promise, this approach currently faces limitations due to the constraints of hardware in the Noisy Intermediate-Scale Quantum (NISQ) era.

    In this study, we introduce and analyze a quantum algorithm for pricing European call options across a broad spectrum of asset models.
    This method transforms a classical approach, which utilizes the Fast Fourier Transform (FFT), into a quantum algorithm, leveraging the efficiency of the Quantum Fourier Transform (QFT).
    Furthermore, we compare this novel algorithm with existing quantum algorithms for option pricing.
\end{abstract}


\section{Introduction}
In the finance sector the pricing of derivatives is an important task.
Derivatives are contracts that promise the buyer a payoff depending on an underlying asset, a simple example is the European Call which pays the buyer the difference between a predefined strike, \(K\), and the value of the underlying at its maturity \(S_T\), if that difference is positive, i.e., \( \max\{S_T - K, 0\} \).
Since the underlying is modeled as a stochastic process, the payoff of this derivative is also stochastic.
Pricing this derivative involves finding the expectation of the payoff under the risk-free measure.
Speeding up this task would be a huge impact quantum computers could bring.
In the literature there are presented different approaches on how to achieve this using a Quantum Computer~\cite{ stamatopoulosOptionPricingUsing2020, wolfQuantumArchitectureSearch2023}.
Both approaches have in common that they mimic classical Monte-Carlo methods, where the potential for improvement comes from the Amplitude Estimation algorithm~\cite{brassardQuantumAmplitudeAmplification2000}.
In this work we will present a different approach that is based on numerically approximating the option price and calculating this approximation via the Discrete Fourier Transform (DFT).
Here the potential speedup stems from the efficiency of the Quantum Fourier Transform (QFT).
This classical part of this method was introduced to price European calls for models where the characteristic function of the log-price is known~\cite{carrOptionValuationUsing1999}.

The remainder of the paper is structured as follows, we will introduce the numerical approximation of the option price in \Cref{sec:numerical_approximation}, and we will see how to perform this approximation on a quantum computer in \Cref{sec:qft_pricing}.
In \Cref{sec:comparison} we will compare the result with other approaches from the literature and give an intuition on error rates in \Cref{sec:error-propagation}.
Finally, we conclude with a short summary and a brief outlook in \Cref{sec:conclusion}.
\section{Numerical Approximation}\label{sec:numerical_approximation}
This section follows the ideas of~\cite{carrOptionValuationUsing1999} to find a numerical approximation of the price of a European Call that can be calculated with the DFT.
The goal is to calculate the fair price of a European call with strike \( \varStrike \in \Rplus\) and maturity \(\varMaturity \in \Rplus \) which is based on an underlying \(\left(S_t\right)_{0 \le t \le \varMaturity}\).
The fair price is then given as the discounted expected payout under the risk-free measure
\begin{equation}\label{eq:option_price}
    C = \e^{-\varInterest\varMaturity} \expectation{\max\left\{S_\varMaturity - \varStrike, 0\right\}}.
\end{equation}
See for example~\cite{kornOptionPricingPortfolio2001} for a more detailed argumentation.

To be able to calculate this expectation we need to assume a model for the underlying.
For this approach we will assume that the characteristic function of the logarithm of the underlying's price is known.
This is the case in many models, the Black-Scholes Model or the Heston Model being two prominent examples or the Variance Gamma model~\cite{madanVarianceGammaModel1990} being a more complex example.

We will calculate the expectation in \Cref{eq:option_price} by expressing it as an integral and approximating this integral numerically.
Note that we replace the underlying's price and the strike with their logarithms and write a lowercase letter instead of a capital one to denote the logarithm.
The Option Price is then given as
\begin{equation}\label{eq:option_price_integral}
    C = \e^{-\varInterest\varMaturity} \int_\varLogStrike^\infty \parens{\e^s - \e^\varLogStrike} f_{s_\varMaturity}\parens{s} \diff s.
\end{equation}
The maximum from \Cref{eq:option_price} is taken care of by starting the integral only from \(\varLogStrike\) instead of \(-\infty\).
If the density \(f_{s_\varMaturity}\) was known analytically, we could stop here, discretize the integral and approximate, but this is not the case for all models we are interested in.

\subsection{Analytic transformation}
First we will transform \Cref{eq:option_price_integral} into an expression that no longer depends on the density, but the characteristic function instead.
As a first step we will express the price in terms of the log-strike and introduce a dampening factor \(\alpha \in \Rplus\) to make the expression square integrable with respect to \(\varLogStrike\).
\begin{equation}\label{eq:option_price_damped}
    \widetilde{C}\parens{\varLogStrike} \coloneqq \e^{\alpha \varLogStrike}C\parens{\varLogStrike} \eqequal{eq:option_price_integral} \e^{\alpha \varLogStrike} \e^{-\varInterest\varMaturity} \int_\varLogStrike^\infty \parens{\e^s - \e^\varLogStrike} f_{s_\varMaturity}\parens{s} \diff s.
\end{equation}
Now we take the Fourier Transform (FT) of \(\widetilde{C}\parens{\varLogStrike}\)
\begin{equation}\label{eq:psi}
    \psi\parens{v} = \int_{-\infty}^\infty \e^{\i v \varLogStrike} \widetilde{C}\parens{\varLogStrike}\diff \varLogStrike.
\end{equation}
Next, we plug in \Cref{eq:option_price_damped} and applying Fubini's Theorem, which we are allowed to do, since the dampening factor guarantees the square integrability.
This leads us to
\begin{displaymath}
    \psi\parens{v} = \int_{-\infty}^\infty \e^{-\varInterest\varMaturity} f_{s_\varMaturity}\parens{s} \int_{-\infty}^s \parens*{\e^s \e^{\varLogStrike \parens{\alpha + \i v}} - \e^{\varLogStrike \parens{1 + \alpha + \i v}}} \diff \varLogStrike \diff s.
\end{displaymath}
The change of order of the integration allows us to explicitly calculate the inner integral
\begin{displaymath}
    \int_{-\infty}^s \parens*{\e^s \e^{\varLogStrike \parens{\alpha + \i v}} - \e^{\varLogStrike \parens{1 + \alpha + \i v}}} = \frac{\e^s \e^{s \parens{\alpha + \i v}}}{\alpha + \i v} - \frac{\e^{s\parens{1 + \alpha + \i v}}}{1 + \alpha + \i v}.
\end{displaymath}
Together with the definition of the characteristic function
\begin{displaymath}
    \Phi\parens{v} \coloneq \int_{-\infty}^\infty \e^{\i v s} f_{s_\varMaturity}\parens{s} \diff s
\end{displaymath}
we can simplify \cref{eq:psi} to
\begin{displaymath}
    \psi\parens{v} = \frac{\e^{-\varInterest\varMaturity} \Phi\parens*{\frac{1+\alpha+\i v}{\i}}}{\parens{\alpha + \i v}\parens{1 + \alpha + \i v}}.
\end{displaymath}
As we require for this approach the characteristic function to be known analytically, we know \(\psi\parens{v}\) analytically as well.

In a last step we apply the inverse FT and reverse the dampening to get an expression for the Option Price
\begin{equation}\label{eq:option_price_transformed}
    C\parens{\varLogStrike} = \e^{-\alpha \varLogStrike}\widetilde{C}\parens{\varLogStrike} = \frac{\e^{-\alpha \varLogStrike}}{2\pi} \int_{-\infty}^{\infty}\e^{-\i v \varLogStrike} \psi\parens{v} \diff v.
\end{equation}

\subsection{Numerical Approximation}
Now that we have an expression for the Price the next step is actually calculating it.
As \cref{eq:option_price_transformed} contains an improper integral, that we cannot calculate directly, we need to find an approximation, i.e., compute the price numerically.
At this point this work deviates from~\cite{carrOptionValuationUsing1999} to make the approximation more suitable for a quantum computer.
We will have to make two approximations in the process of numerically calculating \(C\parens{\varLogStrike}\).
Firstly we have to cut off the integral at some finite point, and secondly we will introduce a discretization error.
The second source of approximation error can be mitigated by employing more involved discretization schemes, like the Simpson Rule, but this is not the focus of this work and when looking at the results later on, not really necessary.

\begin{figure}[!t]
    \centering
    \includegraphics{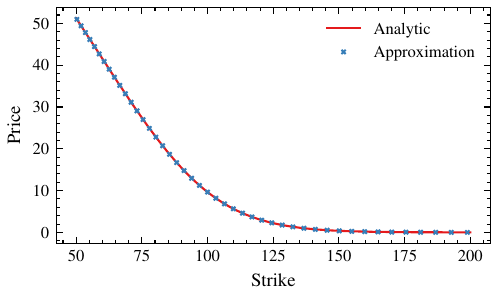}
    \caption{Prices of a European Call in a Black Scholes model with parameters \(S_0 = 100\), \(r = 0.05\), \(\sigma = 0.3\), \(T = 0.5\). The prices are calculated with the discretization from \Cref{eq:dft_price} where \(\alpha = 2.5\), \(N = 2^{10}\), \(\Delta_k = \frac{1}{\sqrt{N}}\) and \(k_0 = \ln 100 - \frac{N \Delta_k}{2}\). We see that the approximation that could be realized with ten qubits is already very}\label{fig:fft_prices}
\end{figure}

\begin{figure}[!t]
    \centering
    \includegraphics{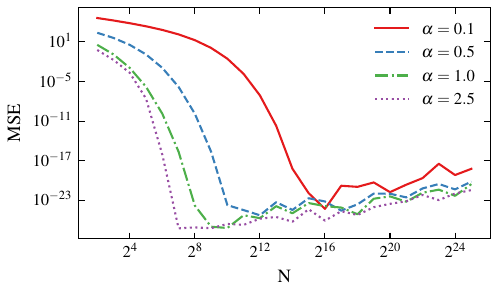}
    \caption{
        Mean squared error between true prices of a European Call and their approximation by \Cref{eq:dft_price} for varying strikes and increasing number of discretization steps in a Black Scholes model with parameters \(S_0 = 100\), \(r = 0.05\), \(\sigma = 0.3\), \(T = 0.5\).
        For all discretizations we fix \(\Delta_k = \frac{1}{\sqrt{N}}\) and \(k_0 = \ln 100 - \frac{N \Delta_k}{2}\).
        We see that the speed of convergence heavily depends on the choice of \(alpha\).
    }\label{fig:fft_convergence}
\end{figure}

We fix a step size \(\Delta_v \in \Rplus\) and the number of discretization steps \(N \in \N\) for the positive and negative part of the integration range each and discretize \Cref{eq:option_price_transformed} as
\begin{displaymath}
    C\parens{\varLogStrike} \approx \frac{\e^{-\alpha \varLogStrike}}{2\pi} \sum_{\indI=-N}^{N-1} \e^{-\i \Delta_v \indI \varLogStrike} \psi_\varMaturity\parens{\Delta_v \indI} \Delta_v.
\end{displaymath}
Now we specify a discretization for the log-strike \(\varLogStrike_\indII \coloneqq \varLowestLogStrike + \Delta_\varLogStrike \indII\) for \(0 \le \indII < 2N \) while we fix the relation between the discretizations of \(v\) and \(\varLogStrike\) to \(\Delta_\varLogStrike \Delta_v = \frac{\pi}{N}\).
With an additional shift of the index this will give us a discretization that fits into the structure needed for the DFT

\begin{equation}\label{eq:dft_price}
    \begin{split}
        C\parens{\varLogStrike_\indII} &\approx \frac{\e^{-\alpha \parens{\varLowestLogStrike + \Delta_\varLogStrike \indII}}}{2\pi} \Delta_v \e^{\i \pi \indII} \sum_{\indI=0}^{2N-1} \e^{-\i \frac{2\pi}{2N} \indI \indII} x_\indI \\
        x_\indI &= \e^{-\i \Delta_v \indI \varLowestLogStrike} \e^{\i \Delta_v N \varLowestLogStrike} \psi\parens{\Delta_v \parens{\indI-N}}.
    \end{split}
\end{equation}
As seen in \Cref{fig:fft_prices,fig:fft_convergence} when choosing a good value for the dampening factor \(\alpha\), even for moderate number of discretization steps the approximation is very accurate.

\section{Pricing with the Quantum Fourier Transform}\label{sec:qft_pricing}
In this section we will see the necessary steps to evaluate the approximation from \Cref{sec:numerical_approximation} on a quantum computer.
The idea is to calculate the sum from \Cref{eq:dft_price} with the QFT and thus achieving a benefit compared to the classical approach.

The QFT is a unitary operation \(QFT_n \in \C^{2^n\times 2^n}\), operating on an \(n\)-qubit system, that transforms a given state \(\Ket{x} = \sum_{\indI=0}^{2^n-1} x_\indI \Ket{\indI}\) into \(\Ket{y} = \sum_{\indII=0}^{2^n-1} y_\indII \Ket{\indII}\) where 
\begin{displaymath}
    y_\indII = \frac{1}{2^{\frac{n}{2}}}\sum_{\indI=0}^{2^n-1} x_\indI \e^{\i \frac{2\pi}{2^n} \indI \indII}.
\end{displaymath}

The inverse QFT is a unitary operation \(QFT_n^{-1} \in \C^{2^n\times 2^n}\), which maps a state \(\Ket{x} = \sum_{\indI=0}^{2^n-1} x_\indI \Ket{\indI}\) to \(\Ket{y} = \sum_{\indII=0}^{2^n-1} y_\indII \Ket{\indII}\) where
\begin{displaymath}
    y_\indII = \frac{1}{2^{\frac{n}{2}}}\sum_{\indI=0}^{2^n-1} x_\indI \e^{-\i \frac{2\pi}{2^n} \indII \indI}.
\end{displaymath}
Note that the conventions for the QFT differ slightly from the classical version, for once the QFT corresponds to the inverse DFT.
Further the factor before the sum is in the QFT setting distributed between the two directions, whereas for the DFT only one direction of the transform has that factor.

Before we can apply the QFT, or to be precise the inverse QFT, we need to prepare the correct state.
As an \(n\)-qubit system allows for \(2^n\) amplitudes we will limit ourselves to discretizations \(N=2^n\) for some \(n \in \N\).
The positive and negative part of the integral will be discretized individually, so in total we need \(n+1\) qubits.
We further have to make sure that the state we want to prepare is actually a valid state.
Thus, we normalize first 
\begin{displaymath}
    \widetilde{x}_\indI = \frac{x_\indI}{\sqrt{\sum\limits_{\indII=0}^{2^{n+1}-1} \abs*{x_\indII}^2}}.
\end{displaymath}
We will call the unitary that prepares this state E
\begin{displaymath}
    E \Ket{0} = \sum_{\indI=0}^{2^{n+1}-1} \widetilde{x}_\indI \Ket{\indI}.
\end{displaymath}
On that state we apply the inverse QFT
\begin{displaymath}
    \Ket{y} = QFT^{-1} E \Ket{0}
\end{displaymath}
where
\begin{align*}
    y_\indII &= \frac{1}{2^{\frac{n+1}{2}}} \sum_{\indI=0}^{2^{n+1}-1} \e^{\i \frac{2\pi}{2^{n+1}} \indII \indI} \widetilde{x}_\indI \\
    &= \frac{1}{2^{\frac{n+1}{2}}} \frac{1}{\sqrt{\sum\limits_{\indI=0}^{2^{n+1}-1} \abs*{x_\indI}^2}}\sum_{\indI=0}^{2^{n+1}-1} \e^{\i \frac{2\pi}{2^{n+1}} \indII \indI} x_\indI.
\end{align*}
Now we can plug this into \Cref{eq:dft_price} to get
\begin{equation}\label{eq:qft_price}
    C\parens{\varLogStrike_\indII} \approx \frac{\e^{-\alpha \parens{\varLowestLogStrike + \Delta_\varLogStrike \indII}}}{2\pi} \Delta_v \e^{\i \pi \indII} \sqrt{\sum_{\indI=0}^{2^{n+1}-1} \abs*{x_\indI}^2} 2^{\frac{n+1}{2}} y_\indII.
\end{equation}

Unfortunately the values \(y_\indII\) we need are encoded in the amplitude of a quantum state, and we do not have direct access to it.
This could be achieved by state tomography, but that is a very inefficient procedure, taking exponentially many measurements in the number of qubits~\cite{jamesMeasurementQubits2001}.
Alternatively we can retrieve the absolute values \(\abs{y_\indII}\) by either Amplitude Estimation or by simply executing many shots.
While the latter is less efficient it estimates all amplitudes in one go and not only the one of a single state.

Knowing that the true price is real and positive and looking at \Cref{eq:qft_price} we see that \(\e^{\i \pi \indII} y_\indII\) needs to be positive and real if the approximation would be without error.
But the approximation is not perfect, thus we introduce another approximation error replacing \(\e^{\i \pi \indII} y_\indII\) by the square root of the observed probability of state \(\Ket{\indII}\).
We will call the observed probabilities \(p_\indII\).
The final approximation is given as
\begin{equation}\label{eq:qft_price_observed}
    C\parens{\varLogStrike_\indII} \approx \frac{\e^{-\alpha \parens{\varLowestLogStrike + \Delta_\varLogStrike \indII}}}{2\pi} \Delta_v  \sqrt{\sum_{\indI=0}^{2^{n+1}-1} \abs*{x_\indI}^2} 2^{\frac{n+1}{2}} \sqrt{p_\indII}.
\end{equation}

\begin{figure}[!t]
    \centering
    \includegraphics{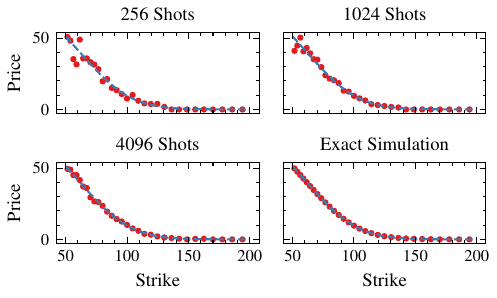}
    \caption{Prices of a European Call for different number of shots on a noise free simulator in a Black Scholes model with parameters \(S_0 = 100\), \(r = 0.05\), \(\sigma = 0.3\), \(T = 0.5\). The prices are calculated with the discretization from \Cref{eq:dft_price} where \(\alpha = 2.5\), \(N = 2^{10}\), \(\Delta_k = \frac{1}{\sqrt{N}}\) and \(k_0 = \ln 100 - \frac{N \Delta_k}{2}\). We see that even for an error free device the error introduced by estimating the probabilities from finite many realizations introduces an error.}\label{fig:qft_prices_shot_noise}
\end{figure}

The quality of the approximation strongly depends on how well the amplitudes are estimated, in \Cref{fig:qft_prices_shot_noise} we can see that this again depends strongly the number of shots used for the approximation, even if the quantum computer is noise free, depicted here by the use of a simulator.

\section{Comparison}\label{sec:comparison}
In this section we will see how this novel approach compares to existing quantum algorithms for pricing options.
The approaches introduced in~\cite{stamatopoulosOptionPricingUsing2020,wolfQuantumArchitectureSearch2023} all have in common, that they start with loading a distribution into a quantum register.
On this distribution they apply a quantum algorithm to encode the options payoff in the amplitude of one target qubit.
This amplitude can then be retrieved either by repeated measurement or amplitude estimation.

For the approach introduced in~\cite{stamatopoulosOptionPricingUsing2020} we will rely on the implementation done by the qiskit-finance \cite{Qiskit} python library, while for our novel QFT approach and the Conditional Parameterized Quantum Circuits (CPQC)~\cite{wolfQuantumArchitectureSearch2023} we will rely on our own implementation.

We will not look at results on actual hardware, as we are interested in comparing the algorithms in general and not how well they cope with current hardware restrictions.

\begin{figure}[!t]
    \includegraphics{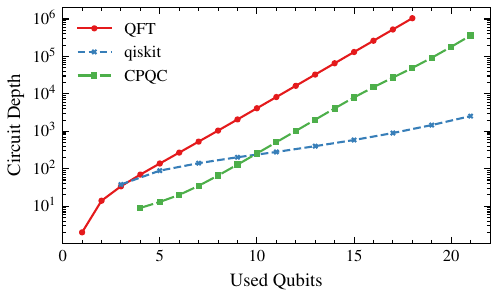}
    \caption{
        The circuit depth for different algorithms depending on the number of used qubits.
        For calculating the depth every circuit is transpiled to consist only of CNOT and U3 gates.
        Note that the circuit depth comes mainly from the state preparation necessary for all algorithms.
        The implementation provided by qiskit-finance and the implementation using CPQCs both use ancillary qubits that make them look better in this comparison, as they do not need to be initialized.
    }\label{fig:algorithm_comparison_qubits_v_depth}
\end{figure}

A common way to analyze quantum algorithms is to look at the scaling of the circuit depth or the number of two qubit gates for an increasing number of qubits. 
We see in \Cref{fig:algorithm_comparison_qubits_v_depth} that this novel approach does not compare well in this metric.

\begin{figure}[!t]
    \centering
    \subfloat[Circuit Depth against Option Price.]{\includegraphics{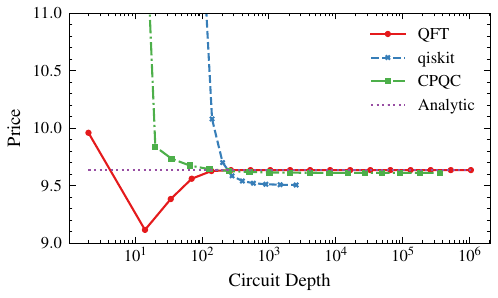}\label{fig:qft_algorithm_comparison_depth_v_price}}
    \hfil
    \subfloat[Circuit Depth against Error.]{\includegraphics{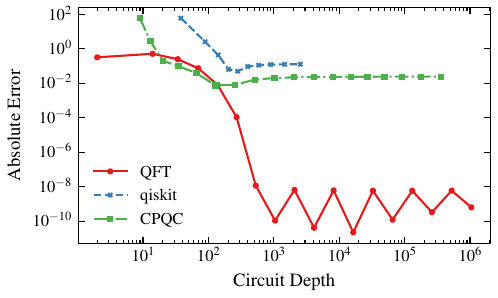}\label{fig:algorithm_comparison_depth_v_error}}
    \hfil
    \subfloat[Used Number of Qubits against Error.]{\includegraphics{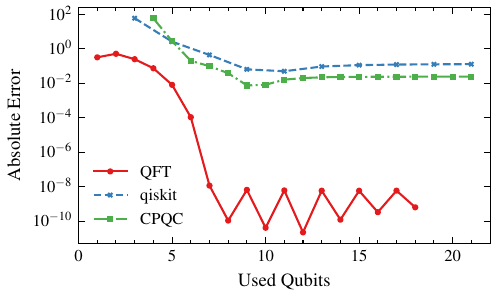}\label{fig:algorithm_comparison_qubits_v_error}}
    \caption{
        Comparison of different quantum algorithms to compute option prices. 
        The priced option is a European Call with strike 100 in a Black Scholes model with parameters \(S_0 = 100\), \(r = 0.05\), \(\sigma = 0.3\), \(T = 0.5\).
        We compare the novel QFT approach with two Monte-Carlo inspired approaches.
        We see in \Cref{fig:qft_algorithm_comparison_depth_v_price} that the qiskit-finance~\cite{Qiskit}implementation is biased, while the other two seem to converge to the true price.
        For finance applications a high precision is important thus we look at that in \Cref{fig:algorithm_comparison_depth_v_error}.
        When comparing the approaches regarding the number of used qubits \Cref{fig:algorithm_comparison_qubits_v_error}, the results look similar.
    }\label{fig:algorithm_comparison}
\end{figure}

That only tells us, that this new approach scales worse for an increasing number of qubits.
But in our setting increasing the number of qubits is solely done to improve the approximation, we don't need a specific number of qubits to be able to calculate real world examples.
This means we should look at the relation between the achieved approximation quality and the quantum resources needed.
As seen in \Cref{fig:algorithm_comparison} in this comparison the QFT based option pricing outperforms the other approaches notably.
This is possible, as the QFT based approach needs fewer qubits for getting good approximations.

A further advantage of the QFT based approach is, that the prices for many strikes can be computed with only one circuit.
The downside is, that it only works for European Calls, and, in extension by the Put Call parity, for European Puts~\cite{hullOptionsFuturesOther2015}.
\section{Error Propagation}\label{sec:error-propagation}
Until now we only looked at idealized error free quantum devices.
In this section we will get an Idea how good a quantum computer would need to perform to get practical results.
To do so we compare \Cref{eq:qft_price} and \Cref{eq:qft_price_observed} the first one being the theoretical approximation making use of the amplitude and the latter being the observable approximation making use of the observed probability.
Let us call the error the quantum computer does \(\epsilon_\indII = \abs{y_\indII - \sqrt{p_\indII}}\) then the absolute difference between the theoretical approximation and the observed approximation is
\begin{equation}\label{eq:error_propagation_factor}
    \mathcal{E}_\indII = \frac{\e^{-\alpha \parens{\varLowestLogStrike + \Delta_\varLogStrike \indII}}}{2\pi} \Delta_v  \sqrt{\sum_{\indI=0}^{2^{n+1}-1} \abs*{x_\indI}^2} 2^{\frac{n+1}{2}} \epsilon_\indII.
\end{equation}
This expression depends on the strike for which we want to calculate the option price, but also on the parameter \(\alpha\), which we can choose freely.
Note that the \(x_j\) also depend on \(\alpha\), and thus it is not straight forward, how exactly the \(\mathcal{E}_\indII\) depend on \(\alpha\).

\begin{figure}[!t]
    \centering
    \subfloat{\includegraphics{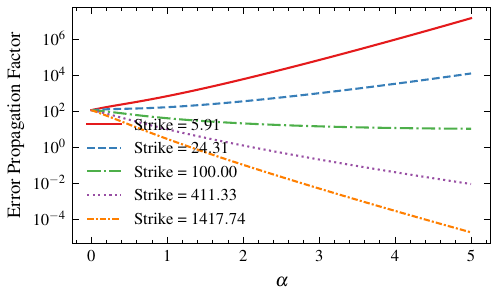}}
    \hfil
    \subfloat{\includegraphics{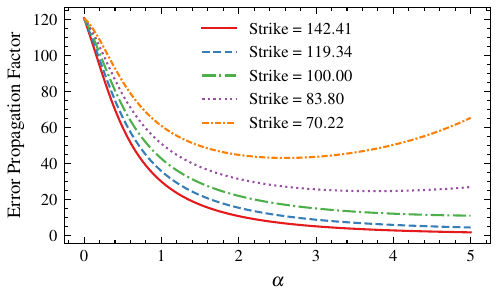}}
    \caption{
        The factor from \Cref{eq:error_propagation_factor} for different strikes and dampening factors \(\alpha\) for a Black Scholes Model with parameters \(S_0 = 100\), \(r = 0.05\), \(\sigma = 0.3\), \(T = 0.5\).
        We can see that there is no one value for \(\alpha\) that fits for all strikes, especially if that strike is far away from the start value of the underlying.}\label{fig:error_propagation}
\end{figure}

In \Cref{fig:error_propagation} we can see that the behavior of this error propagation heavily depends on the strike, especially if the strike is far away from the starting value of the underlying.
This means that the choice of \(\alpha\) could be different, depending on the strike we are interested in.
But as long the strike is not too far from the start value there is a sweet spot for \(\alpha\) around \(2.5\).
There the factor \(\mathcal{E}\) from \cref{eq:error_propagation_factor} is below \(50\) which tells us, that if we, for example, want an approximation precision up to the second digit, the quantum computer needs to approximate the amplitude with an error below \(\frac{0.01}{50} = 0.0002\).
Unfortunately for devices in the NISQ era, this is not achievable.
\section{Conclusion}\label{sec:conclusion}
We have seen a novel approach in using quantum computers for pricing Options that makes use of the QFT.
This new approach performs better than previously introduced algorithms for a wide class of models, but only for European Calls and Puts.
The QFT promises a significant speedup compared to the classical FFT, which is unfortunately in parts counteracted by the expensive state preparation.

As no efficient procedure is currently known for preparing general states, further research could investigate whether there is an inherent structure in the states, needed for this novel approach, that permits more efficient preparation.

\bibliographystyle{IEEEtran}
\bibliography{bibliography}

\end{document}